# Quantum resistant multi-signature scheme with optimal communication round: A Blockchain-based approach

Hamidreza Rahmati[1] , Farhad Rahmati[1]

[1] Department of Mathematics and Computer Science, Amirkabir University of Technology, Tehran, Iran.
hrahmati@aut.ac.ir; frahmati@aut.ac.ir

**Abstract.** Blockchain is a decentralized network to increase trust, integrity, and transparency of transactions. With the exponential growth of transactions in the realm of Blockchain, especially in Bitcoin, Blockchain size increases as all transactions must be stored and verified. In Bitcoin, validating M-of-N transactions involves the necessity of M authentic signatures out of the total N transactions. This procedure is so time-consuming and needs a significant storage capacity. To address these issues, several multi-signature schemes have been proposed, enabling users to interactively generate a common signature on a single message. Recently, some lattice-based multi-signature schemes have been presented to deal with the threats of quantum computers. However, none of them have met all desirable features of multi-signature schemes like aggregate public key, low numbers of communication rounds, or resistant to quantum computers. Within this paper, we present a new multi-signature scheme based on lattices, known as $Razhi-ms$, that has aggregate public key, necessitates solely a single round of communication, and is resistant to quantum computers. In $Razhi-ms$, the aggregate public key size and the final signature size are equal to the public key size and the final signature size of a standard signature respectively, and are independent of the number of signers.

## 1 Introduction

Blockchain [1] is a distributed and decentralized ledger that allows for secure and transparent transaction record-keeping among a network of computers. Blockchain is an immutable ledger of blocks that is used to share and store data in a distributed manner. In contrast to conventional centralized networks, that a unitary entity exercises authority over the data, Blockchain operates on a decentralized peer-to-peer network, making it resistant to manipulation and tampering.

The Blockchain market has steadily grown over the years. The market size of Blockchain technology on a global scale achieved \$11.14 billion in 2022 and is anticipated to expand from \$17.57 billion in 2023 to a substantial \$469.49 billion by the year 2030[1]. In Blockchain, a node after making a transaction and signing it, broadcasts it to the network. Other nodes validate the received transaction and if it is confirmed, they replay it in the network. Miner nodes collect the validated transactions within a block and attach that block to the previous blocks of the blockchain. Standard transactions in the Blockchain only need to send and store a signature between

[1] . https://www.fortunebusinessinsights.com/industry-reports/blockchain-market-100072

transactors, which are known as single-signature transactions. The most prominently talked about Blockchain application is digital currency. Bitcoin [2] is a decentralized digital currency that operates on Blockchain, and allows peer-to-peer transactions to take place on the internet. To increase security and flexibility, Bitcoin Blockchain supports more complex M-of-N transactions [3] to authorize a Bitcoin transaction. In these transactions, M valid signatures from N signatures are required to authorize Bitcoin, and these signatures are stored in the Blockchain. This adds an extra layer of security because even if one private key is compromised or lost, the funds remain secure as long as the remaining private keys are intact. On the other hand, storing M signatures separately, requires more time for verifying and more storage memory. Therefore, minimizing the memory required for storage and the time required for its verification are major concerns.

Multi-signature (MS) schemes are one of the most common tools to solve these problems [4]. A MS scheme is a protocol that allows a set of users to cooperatively generate a common signature $\sigma$ on a unit message $m$, such that the signature verifier is satisfied that each member of the set has taken part in creating the signature. This scheme allows $k$ users to create a compressed signature of approximately the same length as a conventional digital signature. So, the MS can reduce the bandwidth required for $k$ signatures from $\mathcal{O}(k)$ to $\mathcal{O}(1)$. Therefore, compared to the $k$ digital signatures $\{\sigma_1, \dots, \sigma_k\}$, the MS scheme has a lower computational cost, higher verification speed, and requires less storage space. The aggregate public key is the result of combining the public keys of the signers [5] and increase the efficiency of the MS scheme. The aggregate public key size is equal to the public key size of a conventional signature, and the MS scheme can be verified with it. These features are highly useful and suitable for Blockchain-based networks such as Bitcoin [6]. Achieving a set of multiple desirable properties is the primary technical challenge in building a MS scheme. However, none of the existing MS schemes meet all the desirable features. The most important related challenges are:

*Challenge 1*: *Vulnerable to Quantum computers.* The emergence of quantum computers presents a serious security threat to cryptographic systems which are based on classical hard problems such as number theory [7, 8], discrete logarithm [9, 10], and pairing [11, 12]. Peter Shor [13] proved that the mentioned hard problems can be resolved using a quantum computer in polynomial time. Consequently, the cryptographic systems based on those problems break in polynomial time when there is a quantum computer. Although some previous schemes, such as [14], [15], [16], and [17] are resistant to quantum computers, at least two rounds of interaction are required for all of them during the MS generation.

*Challenge 2*: *High number of rounds.* A MS scheme is a protocol that is produced by executing one or more rounds of communication between signers. The computational cost and telecommunication overhead increase as the number of rounds of communication between signers increases. Therefore, it is ideal to provide 1-round MS schemes. Most MS schemes require at least 2 rounds of interaction between signers.

The proposed post-quantum scheme in [18] requires one round of communication between signers, but this scheme does not have the feature of public-key aggregation.

*Challenge 3*: *Lack of aggregate public key.* Among the existing MS schemes, the only scheme presented in the [19] fulfills the features of aggregate public key and resistance to quantum computers with 1-round, but unfortunately it has been proven in the [20] that the said scheme is not unforgeable.

### 1.1 Our Contribution

To address the mentioned challenges, this paper put forwards an optimal lattice-based MS scheme called $Razhi - ms$. The $Razhi - ms$ scheme is the first MS scheme that simultaneously possesses all of the following features.

- *Resistant against Quantum attacks*: As problem structured based on lattice hard problems are resistant against quantum computers, our proposed scheme is designed based on two lattice hard problem called Module Learning With Error (MLWE) and Module Short Integer Solution (MSIS) [21, 22].
- *Optimal communication round*: To achieve the final MS, $Razhi - ms$ requires just one round of communication among signers. The effectiveness of our scheme addresses the challenge described in Challenge 2.
- *Aggregate public key*: $Razhi - ms$ has an aggregated public key and this scheme can be verified with this key. This feature reduces the memory required for key storage and speeds up signature verification. Also, it helps verifiers to verify final signature without the need for public keys of signers. This feature can address Challenge 3 efficiently.
- *Provable security*: We prove that our proposed scheme is secure under the hardness assumptions of MLWE and MSIS.

### 1.2 Organization

The rest of this paper is organized as follows. In Section 2, we review the related work. Section 3 introduces some required preliminaries. We describe the system architecture in Section 4. Our proposed scheme is presented in Section 5. In Section 6, we present concrete parameters for $Razhi - ms$, and prove the correctness and security of $Razhi - ms$ MS scheme. Finally, Section 7 is drawn a conclusion on the proposed scheme.

## 2 Related Works

Blockchain [23] is a digital database or ledger that records transactions over a network of computers, and offers decentralized and secure solutions to various industries and sectors. Blockchain is a distributed ledger technology that facilitates direct transactions between peers, eliminating the requirement for intermediaries, thereby enhancing transparency, security, and efficiency.

Blockchain technology was invented in 2008 [2] to serve as the public transaction ledger for the cryptocurrency Bitcoin, and its use has spread rapidly since that time.

Peer-to-peer transactions are made possible by cryptocurrencies, which provide increased accessibility to people all over the world without the need for middlemen.

The Bitcoin Blockchain encompasses a comprehensive log of all transactions that have ever taken place within the network. With the ongoing processing and inclusion of additional transactions into the Blockchain, its dimensions persistently expand. Based on the statistics published on the site statista.com[2], In 2023, the size of Bitcoin's Blockchain was on the verge of reaching 500 gigabytes, and its database experiences an increase of approximately one gigabyte every few days. Therefore, the memory needed to store all transactions of the Bitcoin network is considered a serious challenge. Additionally, Bitcoin Blockchain enables enhanced security and flexibility by facilitating the implementation of intricate M-of-N transactions for authorizing Bitcoin transactions. Within this framework, a specified number of M valid signatures out of a total of N signatures must be provided to validate the transaction, with all signatures being recorded and stored within the Blockchain. Storing M signatures individually necessitates additional time for verification and storage memory. Consequently, the reduction of memory usage for storage and the acceleration of verification time are significant considerations. MS schemes are one of the ways to address the mentioned challenges.

The concept of MS was first presented by Itakura and Nakamura [4]. In recent years, many MS schemes based on the classical hard problems of number theory [7, 8], discrete logarithm [9, 10] and pairing [11, 12] have been presented. The development of quantum computers poses a serious security threat to cryptographic systems which are based on these classical problems. Peter Shor [13] proved that the mentioned hard problems can be solved using a quantum computer in polynomial time. Consequently, the cryptographic systems based on those problems break in polynomial time when there is a quantum computer.

After presenting Shor's algorithm, a lot of researches have been done to design systems resistant to quantum attacks. Lattice and its hard problems are one of the attractive fields for dealing with quantum threats. Lattice-based MS schemes can be achieved using the hash-and-sign [7] method or the Fiat-Shamir [24-26] technique. In recent years, several lattice-based MS schemes have been presented. Bansarkhani and Strom [17] presented the first lattice-based MS scheme that is secure under the Ring SIS (RSIS) problem. This scheme lacks aggregate public key, requires three rounds of communication among signers, and has a large signature and public key size. Damgard et al. [15] presented a two-round lattice-based MS scheme that did not have the aggregate public key feature. In 2020, Kancel and Dutta [19] presented a MS scheme that is secure under the SIS problem. This scheme is the first MS scheme based on lattice that offers both compression of signatures and aggregate public key. However, Liu et al [20] demonstrated in 2023 that this scheme does not meet the unforgeability property.

[2] . https://www.statista.com/statistics/647523/worldwide-bitcoin-blockchain-size/

In 2022, a lattice-based MS scheme was introduced called $MuSig-L$ [27], which required one round of online communication and one round of offline communication between signers. Yanbo Chen [14] presented the $DualMS$ MS scheme in 2023, which has two rounds of communication between signers and does not require a trapdoor [28, 29], unlike the [15] and [27] schemes. For this reason, the $DualMS$ scheme has a smaller signature size than the [15] scheme and a smaller signature and public key size than $MuSig-L$. In [18], a 1-round MS scheme based on lattices was introduced, but this scheme requires an honest centralized server and a safe communication channel. In this paper, a lattice-based MS scheme based on the Fiat-Shamir technique is presented. This scheme has aggregate public key, and require only one round communication between signers to generate MS.

In Table 1, the $Razhi-ms$ scheme is compared with other lattice-based MS schemes in terms of various features. The only 1-round lattice-based MS schemes are $Razhi-ms$, [19], and [18]. [20] proved that [19] is insecure. The scheme proposed in [18] needs the trustworthy server and a secure communication channel between the server and every signer to keep signers from knowing the trapdoor and forging a MS. The [17] and [15] schemes do not have an aggregate public key. Both [27] and [14] schemes require two rounds of communication between signers to generate MS. Additionally, the only scheme with $UF-ESA$ security property is $Razhi-ms$.

**Table 1.** Comparison of lattice-based MS schemes

| $\boldsymbol{Scheme}$ | $\boldsymbol{Agg}$ | $\boldsymbol{SC}$ | $\boldsymbol{R}$ | $\boldsymbol{UF-CMA}$ | $\boldsymbol{UF-ESA}$ |
|---|---|---|---|---|---|
| [14] | * | * | - | * | - |
| [15] | - | * | - | * | - |
| [17] | - | * | - | * | - |
| [18] | * | - | * | * | - |
| [19] | * | * | * | - | - |
| [27] | * | * | - | * | - |
| $Razhi-ms$ | * | * | * | * | * |

$Agg$: Aggregate public key, $R$: single-round for MS generation, SC: No need for secure channel between server and signers, $UF-CMA$: Unforgeability under chosen message attack, $UF-ESA$: Unforgeability under existed signatures attack.

As can be seen from the table above, our proposed scheme is the only scheme among the lattice-based MS schemes that brings out all the mentioned features. Overall, it can be said that the $Razhi-ms$ scheme has better features than other lattice-based MS schemes.

### 1.3 Technique Overview

The hash-and-sign method [7] or the Fiat-Shamir technique [24-26] can be used to create a MS scheme based on lattices. The scheme presented within this paper is based on the Fiat-Shamir technique. In this technique, the signer's private key, $s$, is multiplied by the challenge $c$ and then the resulting value $sc$ is added with a random vector $y$ to mask the value $sc$.

In this section, for a better understanding of the scheme, a 2-signature scheme is explained. The generalization of this scheme to a MS scheme with an arbitrary number of signers is straightforward. Suppose two entities decide to create a 2-signature scheme on the message $m$. Without loss of generality, we denote the variables and parameters related to the first entity with the index 1, and the variables and parameters related to the second entity with the index 2.

Consider two distinct entities with associated random values $\rho_1'$ and $\rho_2'$. Now, the vectors $\boldsymbol{y_1}$ and $\boldsymbol{y_2}$ are obtained using the $\rho_1'$ and $\rho_2'$ respectively. After generating the $i$-th entity's personal signature $\boldsymbol{z_i} = \boldsymbol{y_i} + \boldsymbol{s_i c_i}$ for $1 \leq i \leq 2$, the random value $\rho_i'$ is encrypted using the other entity's public key and the unsymmetric encryption algorithm. The resulting value is sent to the other entity along with the signature. Therefore, the first entity can calculate the vector $\boldsymbol{y_2}$, and the second entity can calculate the vector $\boldsymbol{y_1}$. So, the first entity can achieve the value $\boldsymbol{s_2 c_2}$, and the second entity can achieve the value $\boldsymbol{s_1 c_1}$. In this way, the value of $\boldsymbol{s_1 c_1} \circledast \boldsymbol{s_2 c_2}$ can be calculated by parts of a 2-signature scheme. Hence, the final common signature of the proposed 2-signature scheme will be $\boldsymbol{z} = (\boldsymbol{y_1} + \boldsymbol{y_2}) + (\boldsymbol{s_1 c_1} \circledast \boldsymbol{s_2 c_2} \bmod \beta)$. Also, the aggregate public key of this scheme is $\mathbf{b} = \boldsymbol{A}(\boldsymbol{s_1 c_1} \circledast \boldsymbol{s_2 c_2} \bmod \beta)$.

# 3 Preliminaries

In this section, some mathematical notations, hard problems and assumptions are presented, and also are introduced the syntax and the security definition of MS scheme.

### 3.1 Notations

Considering the ring $\mathcal{R} = \frac{\mathbb{Z}[x]}{(x^n+1)}$, the quotient ring is shown by $\mathcal{R}_q = \frac{\mathbb{Z}_q[x]}{(x^n+1)}$, where $n$ is a power of 2 and $q$ is a prime number such that $q = 1(\bmod 2n)$. Any element $\mathrm{v} = \sum_{i=0}^{n-1} v_i x^i$ in $\mathcal{R}$ or $\mathcal{R}_q$ is shown by its coefficients vector $(v_0, \dots, v_{n-1})$ that $v_i \in \left[{}^{-(q-1)}/_2, {}^{(q-1)}/_2\right]$. For an even (odd) positive integer $\alpha$, $r' = r \bmod^{\pm} \alpha$ is defined as the distinct element $r'$ within the range $(-\alpha/2) < r' \leq ({}^{\alpha}/_2)$ $\left(-(\alpha-1)/2 \leq r' \leq {}^{(\alpha-1)}/_2\right)$. This element $r'$ is chosen in such a way that it satisfies the congruence relation $r' \equiv r \bmod \alpha$. Matrices and column vectors are shown in bold upper-case and bold lower-case letters respectively, and $\boldsymbol{u}^t$ is the transpose of $\boldsymbol{u}$. The infinity norm is indicated by $\|.\|_\infty$, and for an element $w \in \mathbb{Z}_q$, $\|w\|_\infty$ means $|w \bmod^{\pm} q|$. The infinity norm for $w = w_0 + w_1 X + \cdots + w_{n-1} X^{n-1} \in \mathcal{R}$, means $\|w\|_\infty = \max_i \|w_i\|_\infty$. Also, for $\boldsymbol{w} = (w_1, \dots, w_k) \in \mathcal{R}^k$, we define $\|\boldsymbol{w}\|_\infty = \max_i \|w_i\|_\infty$. For a finite set $X$, $x \leftarrow X$ means that an element $x$ is chosen uniformly at random from the $X$ .$\circledast$ indicates the multiplication of the component by the component of the two vectors. For example, if $\mathbf{v} = (v_1, \dots, v_n)$ and $\mathbf{u} = (u_1, \dots, u_n)$ are two vectors, then $\mathbf{v} \circledast \mathbf{u} = (v_1.u_1, \dots, v_n.u_n)$. The function H is a collision resistant hash ($CRH$) function, and XOF is an eXtendable-Output Function that provides a bitstring of arbitrary length as output. Also, if a statement is true, the boolean operator $[\![\text{statement}]\!]$ returns 1 and returns 0 otherwise.

### 3.2 MS scheme and its security model

**Definition.** A MS scheme is a 4-tuple $\mathrm{MS} = (Setup, KGen, MSign, MSVer)$ where the $Setup$ and $KGen$ algorithms are probabilistic polynomial-time algorithms and the $MSVer$ algorithm is a deterministic polynomial-time algorithm. The signing process $MSign$ is a collaborative protocol among users, where each user executes a polynomial-time signature algorithm to implement this protocol.

- The $\rho \leftarrow Setup(1^\lambda)$ algorithm returns a public parameter $\rho$ with the input of a security parameter $\lambda$.
- Key generation algorithm $(pk_i, sk_i) \leftarrow KGen(\rho)$ for $1 \le i \le n$, which is performed by $i$-th signer independently, and generates a public/private key pair $(pk_i, sk_i)$ for $i$-th signer.
- The $MSign(\rho, sk_i, PK, m)$ signature algorithm, using the public parameter $\rho$, the private key $sk_i$, and the public keys set of users $PK$, generates the joint signature $\sigma$ on the message $m$.
- The signature verification algorithm $b \leftarrow MSVer(\rho, PK, m, \sigma)$ returns $b = 1$ if $\sigma$ is a valid MS for $(PK, m)$, and otherwise $b = 0$.

A MS scheme is required to satisfy the completeness property.

**Definition (Completeness):** Let MS $= (Setup, KGen, MSign, MSVer)$ be a MS scheme. If for each security parameter $\lambda$, each public parameter $\rho$, and each message $m$, always $MSVer(\rho, PK, m, \sigma) = 1$ for a $\sigma \leftarrow MSign(\rho, sk_i, PK, m)$, then MS scheme has the property of completeness.

It should be impossible to forge a valid MS scheme in the presence of at least one honest signer.

**Definition ($UF-CMA$):** A MS scheme MS $= (Setup, KGen, MSign, MSVer)$ is unforgeable under adaptively chosen message attacks, if for each polynomial-time forger $\mathcal{F}$ with known messages and signatures, the probability that $\mathcal{F}$ forge a valid MS on a different unsigned message be negligible.

In this paper, a new security concept -Unforgeability under Existed signature Attack ($UF-ESA$)- is defined. This concept means that if a polynomial-time forger $\mathcal{F}$ has access to all private signatures $\{\sigma_1, \dots, \sigma_k\}$, the $\mathcal{F}$ is still not able to forge the final MS.

**Definition ($UF-ESA$):** A MS scheme MS $= (Setup, KGen, MSign, MSVer)$ is unforgeable under Existed signatures attacks, if for each polynomial-time forger $\mathcal{F}$ having all private signatures $\{\sigma_1, \dots, \sigma_k\}$ on $m$, the probability that $\mathcal{F}$ forge a valid MS on $m$ be negligible.

### 3.3 Hard Assumption

In this section, the standard lattice hard problems, MSIS and MLWE are briefly stated. First, the definition of standard problems SIS and LWE is introduced and then their analogues over ring and module is presented.

**Definition ($\mathbf{SIS}_{q,k,l,\beta}$) [30]:** Consider a uniformly random matrix $\boldsymbol{A} \in \mathbb{Z}_q^{k\times l}$, where its columns are composed $l$ vectors $\boldsymbol{a_i} \in \mathbb{Z}_q^k$. Find a non-zero integer vector $\mathbf{v} \in \mathbb{Z}^l$ such that $\|\mathbf{v}\| \le \beta$ and $\boldsymbol{A}\mathbf{v} = \sum_{i=1}^{k} a_i . v_i = 0 \in \mathbb{Z}_q^k$.

RSIS is obtained in a similar manner to SIS, but instead of using the matrix $\boldsymbol{A} \in \mathbb{Z}_q^{k\times l}$, it is derived by multiplying the ring $\mathcal{R}$ over $\mathbb{Z}$.

**Definition ($\mathbf{RSIS}_{q,l,\beta}$) [30]:** Consider $l$ elements $a_i \in \mathcal{R}_q$ chosen at random with uniform distribution, and a matrix $\boldsymbol{A} \in \mathcal{R}_q^l$ that columns of it are $a_i$. The goal is to identify a non-zero vector $\mathbf{v}$ that holds $\|\mathbf{v}\| \le \beta$ and $\boldsymbol{A}.\mathbf{v} = 0 \in \mathcal{R}_q$.

MSIS is a generalization of the RSIS.

**Definition (MSIS$_{q,k,l,\beta}$) [15]:** Considering a random matrix $\boldsymbol{A} \leftarrow \mathcal{R}_{\mathrm{q}}^{k\times l}$, find the vector $\mathbf{v} \in \mathcal{R}_q^{l+k}$ such that $[\boldsymbol{A}|\mathrm{I}].\mathbf{v} = 0$ and $0 < \|\mathbf{v}\| \leq \beta$.

The algorithm advantage of $\mathcal{A}$ over the MSIS$_{q,k,l,\beta}$ is specified as below [14]:

$$Adv_{\mathrm{MSIS}_{q,k,l,\beta}}(\mathcal{A}) = \Pr\left[\boldsymbol{A}.\mathbf{v} = 0 \wedge 0 < \|\mathbf{v}\| \leq \beta : \boldsymbol{A} \leftarrow \mathcal{R}_q^{k\times l}; \mathbf{v} \leftarrow \mathcal{A}(\boldsymbol{A}) \in \mathcal{R}_q^{k+l}\right]$$

The LWE problem [31] is very similar to the SIS, and these two problems can be meaningfully viewed as dual to one another.

**Definition (LWE Distribution) [30]:** Consider a secret vector $\boldsymbol{s} \in \mathbb{Z}_q^k$. By choosing $\boldsymbol{a} \in \mathbb{Z}_q^k$ randomly, choosing $e \leftarrow \chi$, and outputting $(\boldsymbol{a}, b = \langle \boldsymbol{s}, \boldsymbol{a} \rangle + e \bmod q) \in \mathbb{Z}_q^k \times \mathbb{Z}_q$, the LWE Distribution $A_{s,\chi}$ over $\mathbb{Z}_q^k \times \mathbb{Z}_q$ is sampled.

The LWE problem has two main versions: search, and decision. It has been proven that these two versions are equivalent in terms of hardness [30]. Therefore, only the Decision-LWE (DLWE) is presented here.

**Definition (DLWE$_{q,k,l,\chi}$) [30]:** Provided with $l$ samples $(\boldsymbol{a}_i, b_i) \in \mathbb{Z}_q^k \times \mathbb{Z}_q$ independently, where each of them is followed either: (1) $A_{s,\chi}$ for a uniform and random $\boldsymbol{s} \in \mathbb{Z}_q^k$ that is fixed for every sample, or (2) a uniform distribution, specify which of the two cases it is (with a significant advantage).

Lyubashevsky, Peikert, and Regev [32] introduced Ring LWE (RLWE), the ring-based analogue of LWE.

**Definition (RLWE Distribution) [30]:** Consider a secret $s \in R_q$. By choosing $a \in R_q$ randomly, choosing $e \leftarrow \chi$, and outputting $(a, b = s.a + e \bmod q) \in R_q \times R_q$, the RLWE Distribution $A_{s,\chi}$ over $R_q \times R_q$ is sampled.

**Definition (Decision RLWE$_{q,l,\chi}$) [30]:** Provided with $l$ samples $(a_i, b_i) \in R_q \times R_q$ independently, where each of them is followed either: (1) $A_{s,\chi}$ for a uniform and random $s \in R_q$ that is fixed for every sample, or (2) a uniform distribution, specify which of the two cases it is (with a significant advantage).

MLWE is a generalization of the RLWE.

**Definition (Decision MLWE$_{q,k,l,\chi}$) [15]:** Given a pair $(\boldsymbol{A}, \mathbf{b}) \leftarrow \mathcal{R}_q^{k\times l} \times \mathcal{R}_q^k$ decide whether it was generated uniformly at random from $\mathcal{R}_q^{k\times l} \times \mathcal{R}_q^k$, or it was generated in a way that $\boldsymbol{A} \leftarrow \mathcal{R}_q^{k\times l}$, $\mathbf{s} \leftarrow S_\chi^l \times S_\chi^k$ and $\mathbf{b} \coloneqq [\boldsymbol{A}|\mathrm{I}].\mathbf{s}$.

The algorithm advantage of $\mathcal{A}$ over the MLWE$_{q,k,l,\chi}$ is specified as below [14]:

$$\begin{aligned} Adv_{\mathrm{MLWE}_{q,k,l,\chi}}(\mathcal{A}) \\ &= \Pr\left[\mathcal{A}(\boldsymbol{A}, \mathbf{b}) = 1 : \boldsymbol{A} \leftarrow \mathcal{R}_q^{k\times l}; \mathbf{s} \leftarrow S_\chi^{k+l}; \mathbf{b} \coloneqq [\boldsymbol{A}|\mathrm{I}].\mathbf{s}\right] \\ &- \Pr\left[\mathcal{A}(\boldsymbol{A}, \mathbf{b}) = 1 : (\boldsymbol{A}, \mathbf{b}) \leftarrow \mathcal{R}_q^{k\times l} \times \mathcal{R}_q^k\right]. \end{aligned}$$

## 4 System Architecture

From this point on, for convenience, transactions 2-of-3 are used instead of transactions M-of-N. In other words, if we have a valid 2-signature scheme, we have a valid 2-of-3 transaction. In the following, we present a system for transferring Bitcoin, in which the transfer of Bitcoin is finalized if at least 2 of the 3 entities that own Bitcoins approve it. In other words, a 2-signature scheme should be presented by two of the three entities that own Bitcoin.

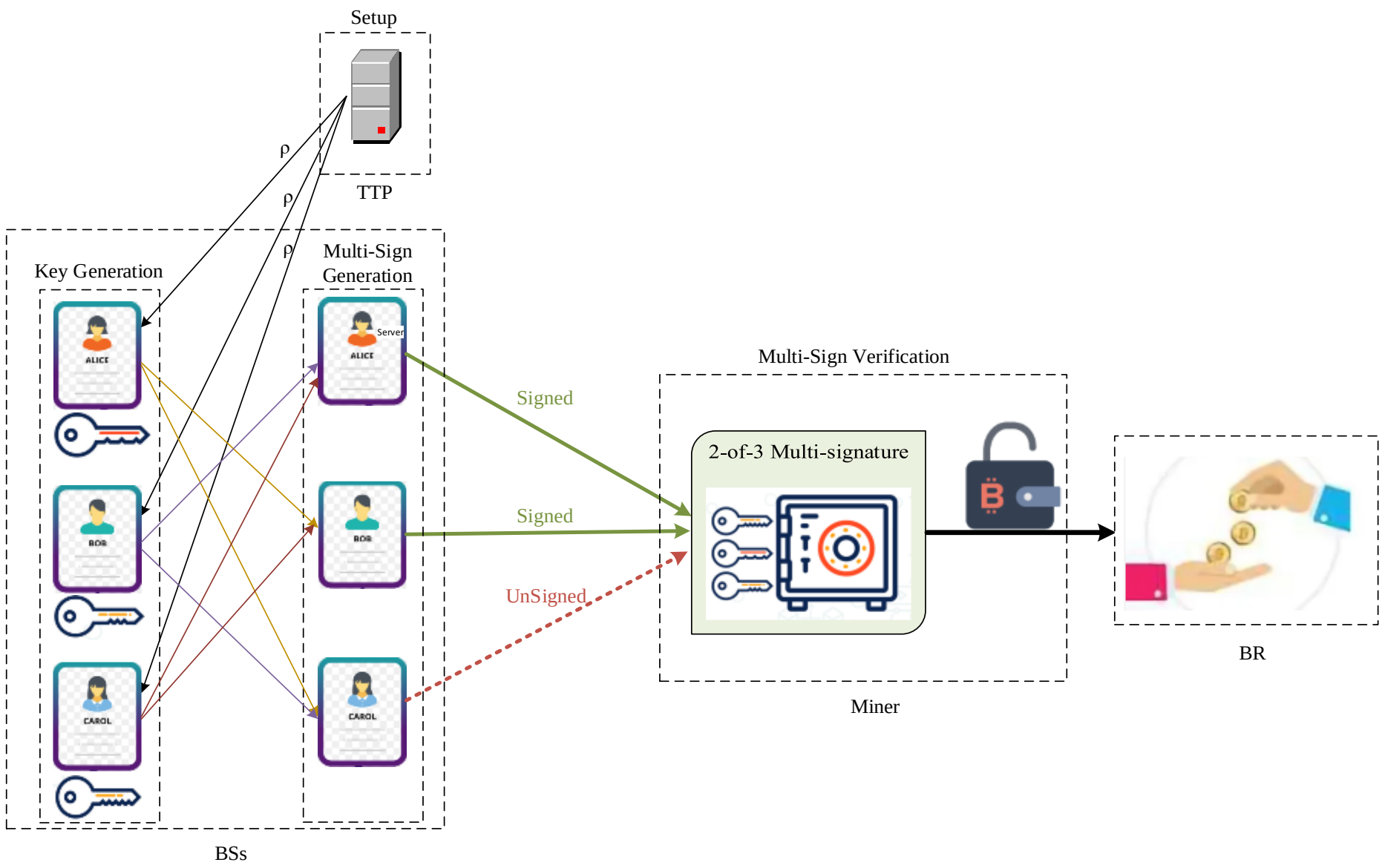

**Fig. 1.** Architecture of proposed 2-of-3 MS scheme

### 4.1 System model

As we have shown in Figure 1, our proposed system consists of four generic entities Trusted Third Party (TTP), Bitcoin Senders (BSs), Bitcoin Recipient (BR), and Miner. Below, the four entities mentioned earlier are described:

- TTP: This entity is responsible for generating Public Parameter ($\rho$), and setting hash function H and XOF.
- BSs: The BSs refer to the individuals or organizations initiating the transaction in the scenario. These entities are the owner of Bitcoin and intend to transfer it to the recipient entity.
- BR: This entity refers to the individual or organization that receives the Bitcoin.
- Miner: Miner is participant in the Bitcoin network responsible for validating and confirming transactions. The Miner gathers transactions, groups them into blocks, and adds them to the blockchain using a process known as mining.

In the following, we give an overview of our 2-of-3 transaction. This transaction is actually a 2-signature scheme. Suppose 3 Bitcoin Senders $BS_1$, $BS_2$, and $BS_3$ define a 2-of-3 transaction. So, a Bitcoin address is associated with three different private keys, and any two out of those three keys are required to authorize a transaction. As shown in Figure 1, our proposed system consists of 4 different phases Setup, Key Generation, Multi-Sign Generation, and Multi-Sign Verification described below:

1. Setup: This phase is implemented by the TTP and generates 256 random bits by taking the security parameter $\lambda$. Also, a hash function $\mathrm{H}: \{0,1\}^* \rightarrow \{0,1\}^{256}$ and a XOF: $\{0,1\}^{256} \rightarrow \{0,1\}^*$ are also set in this phase.
2. Key Generation: This phase is managed by the BSs, and consists of two stages.
   - In this stage, the BSs generate three distinct private keys, and their corresponding public keys. BSs publish their public keys to the other parties and keep the secret keys confidential by themselves.
   - BSs generate the Bitcoin address from the combination of these three public keys, and they agree that the Bitcoin will be transferred to the BR if at least two BSs have confirmed it.
3. Multi-Sign Generation: This phase is executed by the BSs. In order to utilize Bitcoin from this 2-signature address, it is imperative to generate a transaction and authenticate it by appending signatures from a minimum of two out of the three private keys. After generating a 2-signature scheme, the BSs send it on the network.
4. Multi-Sign Verification: This phase is managed by the Miner. Upon receipt of the 2-signature scheme, should the Miner authenticate that 2 of the 3 Bitcoin owner entities have endorsed the Bitcoin transfer, the transaction will be validated. So, the Bitcoin is transferred to the BR. Now, the Miner puts this transaction in a new block and adds it to the Blockchain.

**Table 2.** Notations

| Notation | Description |
|---|---|
| MS | Multi-Signature |
| $\circledast$ | Multiplication of the component by the component of the two vectors |
| $\lambda$ | Security Parameter |
| $\rho$ | Public Parameter |
| $B_\tau$ | The set of elements of $\mathcal{R}$ that have $\tau$ $\pm 1$ and the rest 0 |
| $\lVert . \rVert_\infty$ | Infinity norm |
| $[\![\text{statement}]\!]$ | Boolean operator that returns 1 if statement is true and returns 0 otherwise |
| H | Collision Resistant Hash |
| XOF | eXtendable-Output Function |
| TTP | Trusted Third Party |
| BS | Bitcoin Sender |
| BR | Bitcoin Recipient |
| $\sigma$ | Final MS |
| $\sigma_i$ | Individual signature |
| $\mathcal{F}$ | Forger |
| $\mathrm{HighBits}_q$ | The high-order bits of a vector |
| $\mathrm{LowBits}_q$ | The low-order bits of a vector |

### 4.2 Threat model

TTP is an honest entity and all other entities trust it. BSs that create a 2-of-3 transaction trust each other, and at least 2 of them must cooperate to create a 2-of-3 transaction. No BS alone can create a valid 2-of-3 transaction. Also, no BS can collude with other BSs (except for the BSs who have created the 2-of-3 transaction together) and produce a valid 2-of-3 transaction. The Miner is assumed to be honest, and it always executes the given protocols correctly. Miner does not collude with BSs and does not confirm invalid transactions. It is assumed that the BR is malicious, although it cannot earn any Bitcoins without the transaction being confirmed by the Miner. Also, if a malicious attacker obtains the private key of one of the BSs, it cannot generate any valid 2-of-3 transaction.

## 5 The Proposed scheme

This section presents the proposed 1-round lattice-based MS scheme $Razhi - ms$. As mentioned in section 4.1, our proposed scheme consists of 4 different phases Setup, Key Generation, Multi-Sign Generation, and Multi-Sign Verification. Table 2 describes the notations employed in this section.

### 5.1 Scheme description

Suppose $\{\mathrm{BS}_i\}_{i=1}^{n}$ decide to sign a joint message $m$, that $m$ is the transfer of Bitcoin to the BR. Without losing generality, the variables and values related to $\mathrm{BS}_i$ are shown with index $i$.

**Setup.** In this phase, the TTP selects a security parameter $\lambda$. Then, it executes $\rho \leftarrow Setup(1^{\lambda})$ to generates public parameter $\rho \leftarrow \{0,1\}^{256}$ uniformly, and publishes $\rho$ to the BSs (Figure 2). Also, TTP executes the standard function SHA-3 [33] to generate the hash function H and the XOF.

| $\rho \leftarrow \boldsymbol{Setup}(1^{\lambda})$ |
|---|
| 1. $\rho \leftarrow \{0,1\}^{256}$<br>2. XOF ←SHA-3<br>3. H ←SHA-3<br>4. $Return\ (\rho, \mathrm{XOF}, \mathrm{H})$ |

**Fig. 2.** The pseudo-code for $Razhi - ms$ Setup

**Key Generation.** This phase is managed by the BSs. Consider a group of Bitcoin Senders denoted as $\mathrm{BS}_i$ for $1 \leq i \leq n$, where $n$ denotes number of $\mathrm{BS}_i$. In this phase, $\{\mathrm{BS}_i\}_{i=1}^{n}$ generates its own public/private key by taking the input $\rho$. First, $\{\mathrm{BS}_i\}_{i=1}^{n}$ generate the random uniform matrix $\boldsymbol{A} \in \mathcal{R}_q^{l\times l}$ by running the XOF function. $\{\mathrm{BS}_i\}_{i=1}^{n}$ also sample $(\boldsymbol{s_i}, \boldsymbol{e_i}) \leftarrow S_\eta^l \times S_\eta^k$, $(\boldsymbol{r_{ij}}, \boldsymbol{e'_{ij}}) \leftarrow S_\eta^l \times S_\eta^l$ and $e''_{ij} \leftarrow S_\eta$ uniformly.

| $(pk_i, sk_i) \leftarrow \boldsymbol{Key\ Generation(\rho)}$: |
|---|
| 1. $\boldsymbol{A} \in \mathcal{R}_q^{l\times l} \coloneqq \mathrm{XOF}(\rho)$<br>2. $(\boldsymbol{s_i}, \boldsymbol{e_i}) \leftarrow S_\eta^l \times S_\eta^l$<br>3. $\mathbf{b}_i = \boldsymbol{A}\boldsymbol{s_i} + \boldsymbol{e_i}$<br>4. $for\ i = 2 : n$<br>5. $\{(\boldsymbol{r_{ij}}, \boldsymbol{e'_{ij}}) \leftarrow S_\eta^l \times S_\eta^l\ and\ \ e''_{ij} \leftarrow S_\eta\}$<br>6. $Return\ \left(pk_i = (\rho, \mathbf{b}_i), sk_i = (\boldsymbol{s_i}, \boldsymbol{e_i}, \boldsymbol{r_{ij}}, \boldsymbol{e'_{ij}}, e''_{ij})\right)$ |

**Fig. 3.** The pseudo-code for $Razhi - ms$ Key Generation

Finally, they output the public/private key pair for $1 \le i \le n$ as $(pk_i = (\rho, \mathbf{b}_i = \boldsymbol{A}\boldsymbol{s}_i + \boldsymbol{e}_i),\ sk_i = (\boldsymbol{s}_i, \boldsymbol{e}_i, \boldsymbol{r}_i, \boldsymbol{e}'_i, e''_i))$ (Figure 3).

$\boldsymbol{\sigma} \leftarrow \boldsymbol{Multi - Sign\ Generation}(\boldsymbol{\rho},\ \boldsymbol{m},\ \{\boldsymbol{\sigma_{ij}}\}_{\boldsymbol{i,j=1}}^{\boldsymbol{n}})$:

$\rho'_1 \leftarrow \{0,1\}^{256}$
$k \leftarrow 0$
$\boldsymbol{z} \leftarrow \perp$
$while\ \boldsymbol{z} \leftarrow \perp\ \ do$
$\quad \boldsymbol{y_1} \in S^l_{\gamma_1 - 1} \coloneqq \mathrm{XOF}(\rho'_1)$
$\quad \mathbf{w_1} = \boldsymbol{A}\boldsymbol{y_1}$
$\quad \mathbf{w'_1} = \mathrm{HighBits_q}(\mathbf{w_1}, 2\gamma_2)$
$\quad \boldsymbol{c_1} \in B_{41} \coloneqq \mathrm{H}(m||\mathbf{w'_1})$
$\quad \boldsymbol{z_1} = \boldsymbol{y_1} + \boldsymbol{c_1}\boldsymbol{s_1}$
$\quad \boldsymbol{r}_0 \leftarrow \mathrm{LowBits_q}(\mathbf{w} - \boldsymbol{c_1}\boldsymbol{e_1})$
$\quad if\ \{\|\boldsymbol{z_1}\|_\infty \ge \gamma_1 - \beta\ \ or\ \ \|\boldsymbol{r_0}\|_\infty \ge \gamma_2 - \beta\}$
$\qquad then\ \ \boldsymbol{z} \leftarrow \perp$
$\quad end\ if$
$\quad k \leftarrow k + 1$
$end\ while$
$For\ i = 2 : n\ \{$
$\quad \boldsymbol{u_{1i}} = \boldsymbol{A}\boldsymbol{r_1} + \boldsymbol{e'_1}\ \ \&\ \ \boldsymbol{v_{1i}} = \mathbf{b}_i\boldsymbol{r_1} + \boldsymbol{e''_1} + \rho'_1\}$
$\quad Return\ \sigma_{1i} = (\boldsymbol{z_1}, \boldsymbol{c_1}, \boldsymbol{u_{1i}}, \boldsymbol{v_{1i}})\ \ for\ 2 \le i \le n$

$\xrightarrow{\sigma_{12}\ for\ \mathrm{BS}_2}$
$\vdots$
$\xrightarrow{\sigma_{1n}\ for\ \mathrm{BS}_n}$
$\xleftarrow{\sigma_{21}\ from\ BS_2}$
$\vdots$
$\xleftarrow{\sigma_{n1}\ from\ \mathrm{BS}_n}$
$\boldsymbol{MS}(\boldsymbol{m}, \boldsymbol{\sigma_i})$:
$For\ (\ i = 2{:}\,n)\ \{$
$\quad \boldsymbol{A}\boldsymbol{z_i} - \boldsymbol{c_i}\mathbf{b}_i = \boldsymbol{A}\boldsymbol{y_i} - \boldsymbol{c_i}\boldsymbol{e_i}$
$\quad \mathbf{w}''_i = \mathrm{HighBits_q}(\boldsymbol{A}\boldsymbol{y_i} - \boldsymbol{c_i}\boldsymbol{e_i})$
$\quad if\ \ \{\|\boldsymbol{z_i}\|_\infty < \gamma_1 - \beta\ \ \&\ \boldsymbol{c_i} = \mathrm{H}(m||\mathbf{w}''_i), then$
$\qquad \boldsymbol{v_{i1}} - \boldsymbol{u_{i1}}\boldsymbol{s_1} = \rho'_i$
$\qquad \boldsymbol{y_i} \coloneqq \mathrm{XOF}(\rho'_i)$
$\qquad \boldsymbol{z_i} - \boldsymbol{y_i} = \boldsymbol{c_i}\boldsymbol{s_i}$
$\quad Else\ \perp\ \}$
$\boldsymbol{y} = (\sum_{i=1}^{n} \boldsymbol{y_i}) \bmod \gamma_1$
$\boldsymbol{c} = \mathrm{H}(m||\boldsymbol{A}\boldsymbol{y})$
$\boldsymbol{z} = \boldsymbol{y} + (\circledast_{i=1}^{n} \boldsymbol{c_i}\boldsymbol{s_i} \bmod \beta)$
$\mathbf{b} = \boldsymbol{A}(\circledast_{i=1}^{n} \boldsymbol{c_i}\boldsymbol{s_i} \bmod \beta)$
$Return\ \sigma = (\boldsymbol{z}, \boldsymbol{c}, \mathbf{b})$

**Fig. 4.** The pseudo-code for $Razhi - ms$ Multi-Sign Generation

**Multi-Sign Generation.** This phase is executed by BSs and includes of two steps. The step one is to generate the personal signature of each $\{\mathrm{BS}_i\}_{i=1}^{n}$ on the message $m$.

For a better understanding of the work process, we explain how to generate a signature by $\mathrm{BS}_1$. The method of signature generation by other $\{\mathrm{BS}_i\}_{i=2}^{n}$ is exactly similar to this process. First, $\mathrm{BS}_1$ generates the vector $\boldsymbol{y_1} \in S_{\gamma_1-1}^{l}$ by running the XOF function and with random value input $\rho_1'$, and then by multiplying the matrix $\boldsymbol{A}$ by the vector $\boldsymbol{y_1}$, it obtains the value of $\mathbf{w_1}$. Next, $\mathrm{BS}_1$ assigns the value of $\mathbf{w_1'}$ to the high-order bits of the $\mathbf{w_1}$ (Figure 5). Now $\mathrm{BS}_1$ products the challenge $\boldsymbol{c_1} \in B_{41}$ by executing hash function $\mathrm{H}(m||\mathbf{w_1'})$, and finally generates its personal signature as $\boldsymbol{z_1} = \boldsymbol{y_1} + \boldsymbol{c_1 s_1}$. Also, $\rho_1'$ is encrypted as $\boldsymbol{v_{1i}} = \mathbf{b_i}\boldsymbol{r_1} + \boldsymbol{e_1''} + \rho_1^{'}$ by public key of $\{\mathrm{BS}_i\}_{i=2}^{n}$. In order to the other parties to perform the decryption operation correctly, $\boldsymbol{u_{1i}} = \boldsymbol{A r_1} + \boldsymbol{e_1'}$ is calculated too. Finally, the output is sent to $\mathrm{BS}_i$ as $\sigma_1 = (\boldsymbol{z_1}, \boldsymbol{c_1}, \boldsymbol{u_{1i}}, \boldsymbol{v_{1i}})$. All the above operations are executed in parallel by $\{\mathrm{BS}_i\}_{i=2}^{n}$ and the output $\sigma_i = (\boldsymbol{z_i}, \boldsymbol{c_i}, \boldsymbol{u_{i1}}, \boldsymbol{v_{i1}})$ is sent to $\mathrm{BS}_1$.

In second step, $\mathrm{BS}_1$ after receiving $\sigma_i = (\boldsymbol{z_i}, \boldsymbol{c_i}, \boldsymbol{u_{i1}}, \boldsymbol{v_{i1}})$, first verify that $\boldsymbol{z_i}$ was actually generated by $\{\mathrm{BS}_i\}_{i=2}^{n}$. For this reason, first $\boldsymbol{A z_i} - \boldsymbol{c_i}\mathbf{b_i} = \boldsymbol{A y_i}$ is formed and then it calculates $\mathrm{H}(m||\boldsymbol{A y_i})$, and compares its value with $\boldsymbol{c_i}$. If the above two values are equal, it is ensured that the signature $\boldsymbol{z_i}$ was generated by $\mathrm{BS}_i$ on the message $m$. Now, $\mathrm{BS}_1$ by using its private key and forming $\boldsymbol{v_{i1}} - \boldsymbol{u_{i1} s_1}$, obtains the value of $\rho_i{}'$ and subsequently $\boldsymbol{y_i}$ and $\boldsymbol{c_i s_i}$. The other $\{\mathrm{BS}_i\}_{i=2}^{n}$ also perform all the above tasks by getting $\sigma_1 = (\boldsymbol{z_1}, \boldsymbol{c_1}, \boldsymbol{u_{1i}}, \boldsymbol{v_{1i}})$ and obtaines the values of $\boldsymbol{y_1}$ and $\boldsymbol{c_1 s_1}$. At this step, $\{\mathrm{BS}_i\}_{i=1}^{n}$ arrive at the common MS scheme $\boldsymbol{z} = \boldsymbol{y} + (\circledast_{i=1}^{n} \boldsymbol{c_i s_i} \bmod \beta)$ by forming $\boldsymbol{y} = (\sum_{i=1}^{n} \boldsymbol{y_i}) \bmod \gamma_1$ and then $\boldsymbol{c} = \mathrm{H}(m||\boldsymbol{A y})$. Also, their aggregate public key will be $\mathbf{b} = \boldsymbol{A}(\circledast_{i=1}^{n} \boldsymbol{c_i s_i} \bmod \beta)$. Therefore, the final output is $\sigma = (\boldsymbol{z}, \boldsymbol{c}, \mathbf{b})$ (Figure 4).

In the Figure 5, the auxiliary algorithms required in the MS generation section and its verification are given [34].

| | |
|---|---|
| $\mathrm{Decompose}_q(r, \alpha)$ | $\mathrm{HighBits}_q(r, \alpha)$ |
| ------------------------ | --------------------- |
| $r \coloneqq r \bmod^{+} q$ | $(r_1, r_0) \coloneqq \mathrm{Decompose}_q(r, \alpha)$ |
| **if** $r - r_0 = q - 1$ | **return** $r_1$ |
| **then** $r_1 \coloneqq 0;\ r_0 := r_0 - 1$ | |
| **else** $r_1 \coloneqq {(r - r_0)}/{\alpha}$ | $\mathrm{LowBits}_q(r, \alpha)$ |
| **return** $(r_1, r_0)$ | --------------------- |
| | $(r_1, r_0) \coloneqq \mathrm{Decompose}_q(r, \alpha)$ |
| | **return** $r_0$ |

**Fig. 5.** auxiliary algorithms in proposed MS scheme

**Multi-Sign Verification.** This phase is operated by Miner. Upon receiving the pair $(m, \sigma)$, the Miner first reaches the $\boldsymbol{Ay}$ value using the $\boldsymbol{Az} - \mathbf{b}$, then compares the output of the $\mathrm{H}(m||\boldsymbol{Ay})$ function with challenge $\boldsymbol{c}$. If the above two values are equal, the MS is accepted, otherwise the MS scheme is rejected (Figure 6).

| $\{0,1\} \leftarrow \boldsymbol{Multi - Sign\ Verification}(\boldsymbol{m}, \boldsymbol{\sigma})$: |
|---|
| 1. $\mathbf{w} = \boldsymbol{Az} - \mathbf{b}$ |
| 2. $return\ [\![\boldsymbol{c} = \mathrm{H}(m\Vert\mathbf{w}]\!]$ |

**Fig. 6.** The pseudo-code for Multi-Sign Verification of proposed scheme

## 6 Concrete Parameters, Completeness and security

In the Table 3, we present the set of proposed concrete parameters for the $Razhi - ms$ scheme with the aim of classic 128-bit security.

**Table 3.** Proposed parameters for $Razhi - ms$ scheme

| Parameter | Description | Value |
|---|---|---|
| $n$ | Degree of $f(x)$ that is a power of 2 | 256 |
| $q$ | Prime modulus | 8397313 |
| $(k,l)$ | Dimensions of matrix $A$ | (4,4) |
| $\gamma_1$ | Coefficient range of y | $2^{17}$ |
| $\gamma_2$ | low-order rounding range | 63616 |
| $\tau$ | # of $\pm 1's$ in polynomial $c$ | 41 |
| $\eta$ | Private key range | 5 |
| $\beta$ | $\tau.\eta$ | 205 |
| $\lambda$ | Security parameter | 128 |

In the Table 4, the size of the final signature (|Sig|) and aggregate public key (|APK|), and the level of security of the $Razhi - ms$ scheme is compared with other two lattice-based MS schemes. The security estimator of CRYSTALS[3] has been utilized to estimate the hardness of the MSIS and MLWE.

**Table 4.** Comparison of lattice-based MS schemes

| Scheme | \|APK\| | \|Sig\| | Quantum Core-SVP | MSIS | MLWE |
|---|---|---|---|---|---|
| $\boldsymbol{Musig - L}$ | 26000 | 98000 | 125 | 125 | >700 |
| $\boldsymbol{Dual - MS}$ | 6900 | 19720 | 130 | 130 | 130 |
| $\boldsymbol{Razhi - ms}$ | 3072 | 2214 | 129 | 133 | 129 |

### 6.1 Completeness

**Theorem 1.** Our proposed MS scheme is completeness

**Proof.** To prove the completeness of our scheme, it is first shown that the individual signature produced by each of the BSs is correctness. We show this issue only for the $\text{BS}_1$. Because $\mathbf{w_1} = \boldsymbol{Ay_1}$ and $\mathbf{b_1} = \boldsymbol{As_1} + \boldsymbol{e_1}$, therefore

$$\boldsymbol{Az_1} - \boldsymbol{c_1}\mathbf{b_1} = \boldsymbol{Ay_1} - \boldsymbol{c_1 e_1}$$

Because $\|\boldsymbol{c_1 e_1}\|_\infty \leq \tau.\eta = \beta$ and $r_0 = \text{LowBits}_\text{q}(\mathbf{w_1} - \boldsymbol{c_1 e_1}, 2\gamma_2) < \gamma_2 - \beta$, according to Lemma 2 from [34],

$$\mathbf{w_1''} = \text{HighBits}_\text{q}(\mathbf{w_1} - \boldsymbol{c_1 e_1}, 2\gamma_2) = \text{HighBits}_\text{q}(\mathbf{w_1}, 2\gamma_2) = \mathbf{w_1'}$$

Therefore, $\boldsymbol{c_1} = \text{H}(m||\mathbf{w_1''})$ and hence the correctness of the signature is confirmed. So, the individual signature produced by each of the BSs is correctness. Now, the completeness of $Razhi - ms$ MS scheme is proved for two BSs. The generalization of this proof for arbitrary number of BSs is straightforward and we omit it in this section.

Because $\mathbf{b} = \boldsymbol{A}(\boldsymbol{c_1 s_1} \circledast \boldsymbol{c_2 s_2} \bmod \beta)$ and $\boldsymbol{z} = \boldsymbol{y} + (\boldsymbol{c_1 s_1} \circledast \boldsymbol{c_2 s_2} \bmod \beta)$, so

[3] . https://github.com/pq-crystals/security-estimates

$$\mathbf{w} = \boldsymbol{Az} - \mathbf{b} = \boldsymbol{A}(\boldsymbol{y} + (\boldsymbol{c_1 s_1} \circledast \boldsymbol{c_2 s_2}) \bmod \beta) - \mathbf{b}$$
$$= \boldsymbol{Ay} + \boldsymbol{A}(\boldsymbol{c_1 s_1} \circledast \boldsymbol{c_2 s_2}) \bmod \beta - \boldsymbol{A}(\boldsymbol{c_1 s_1} \circledast \boldsymbol{c_2 s_2}) \bmod \beta = \boldsymbol{Ay}$$

Therefore, in case of producing a valid signature, both the BSs and the Miner will reach a common value $\boldsymbol{Ay}$ and then a common $\boldsymbol{c}$. Hence, the signature is well verified. So $Razhi - ms$ scheme has the feature of completeness. ■

### 6.2 Security

**Theorem 2.** $Razhi - ms$ proposed scheme has $UF - CMA$ security under MSIS and MLWE assumptions.

**Proof.** Suppose attacker $\mathcal{A}$ pretends all the keys $\{2, \dots, k\}$ and finally communicates with the honest signer $\{1\}$. Also suppose that the attacker $\mathcal{A}$ convinces the honest signer $\{1\}$ that the signatures $\{\sigma_2, \dots, \sigma_k\}$ are the real signatures of people $\{2, \dots, k\}$ on message $m$. The honest signer sends its signature $(\boldsymbol{z_1}, \boldsymbol{c_1}, \boldsymbol{u_{1i}}, \boldsymbol{v_{1i}})$ to the attacker $\mathcal{A}$. Assuming that the MSIS and MLWE problems are hard, the attacker cannot obtain any information from $(\boldsymbol{z_1}, \boldsymbol{c_1}, \boldsymbol{u_{1i}}, \boldsymbol{v_{1i}})$. According to Theorem 3.3 in [35], it suffices to prove that an attacker who has access to a random $(\boldsymbol{A}, \mathbf{b})$ cannot produce a pair of message and signature $(m, (\boldsymbol{z}, \boldsymbol{c}, \mathbf{b}))$ that

- $||\boldsymbol{z}|| < \gamma_1 - \beta$
- $\mathrm{H}(m||\boldsymbol{Ay}) = \boldsymbol{c}$

In other words, we need to prove that it is hard for a quantum attacker to generate a pair of message and signature $(m, (\boldsymbol{z}, \boldsymbol{c}, \mathbf{b}))$ under the following conditions:

- $||\boldsymbol{z}|| < \gamma_1 - \beta$
- $||\boldsymbol{c}||_\infty = 1$
- $m \in \{0,1\}^*$
- $\mathrm{H}(m||\boldsymbol{Ay}) = \boldsymbol{c}$

The H function is a hash function that is entirely separate from the algebraic structure of its inputs. Therefore, $m$ can be assumed to be constant. With the same reasoning, the only way to get the answer is to choose a $\boldsymbol{\omega} = \boldsymbol{Ay}$ and then calculate $\mathrm{H}(m||\boldsymbol{\omega}) = \boldsymbol{c}$ and find $\boldsymbol{z}$ such that

$$\boldsymbol{Az} = \boldsymbol{\omega} + \mathbf{b}$$

Finding such a $z$ that holds for $\mathbf{b}' = \boldsymbol{\omega} + \mathbf{b}$ at $\boldsymbol{Az} = \mathbf{b}'$ is equivalent to the difficulty of the MSIS problem. ■

**Theorem 3.** $Razhi - ms$ has $UF - ESA$ security under MSIS and MLWE assumptions.

**Proof.** Suppose the attacker $\mathcal{A}$ has access to the signature oracle and obtains the signatures $\{\sigma_1, \dots, \sigma_k\}$ that $\sigma_i = (\boldsymbol{z_i} = \boldsymbol{y_i} + \boldsymbol{c_i s_i}, \boldsymbol{c_i})$ on the fixed message $m$ from the signature oracle. The attacker has two ways to forge $Razhi - ms$:

1. He should find the signature $\sigma = (\boldsymbol{z}, \boldsymbol{c}, \mathbf{b})$ on $m$ that

- $||\boldsymbol{z}|| < \gamma_1 - \beta$
- $||\boldsymbol{c}||_\infty = 1$

- $\boldsymbol{Ay} = \boldsymbol{A}(\boldsymbol{y_1} + \cdots + \boldsymbol{y_k})$
- $\mathrm{H}(m||\boldsymbol{Ay}) = \boldsymbol{c}$

Obviously, the conditions above are more difficult than the conditions for proving the security of $UF - CMA$ in Theorem 2, so the attacker is not able to forge the $Razhi - ms$ in this way.

2. The attacker must be able to separate the elements of $\boldsymbol{z_i}$ i.e., $\boldsymbol{y_i}$ and $\boldsymbol{c_i s_i}$. For this, the attacker must be able to recover $\rho_i'$. Recovering $\rho_i'$ from $\boldsymbol{v_{1i}}$ and $\boldsymbol{u_{1i}}$ is equivalent to solving the MLWE hard problem. So, the attacker is not able to forge $Razhi - ms$ in this way either.

Therefore $Razhi - ms$ scheme has $UF - ESA$ security. ∎

## 7. Conclusion

In this paper, the $Razhi - ms$ MS scheme along with concrete parameters is presented. $Razhi - ms$ is the first lattice-based MS scheme with $UF - CMA$ security and aggregate public key property, that requires only one round of communication between signers. Unlike most MS schemes where the final signature $\sigma$ is derived from the sum of individual signatures $\sigma_i$, the $Razhi - ms$ scheme is designed in a way that uses two addition and multiplication operators to generate the final signature. This means that if an attacker has access to all of $\sigma_i$'s signatures, he still cannot forge the final signature of $\sigma$. Therefore, unlike most MS schemes, the $Razhi - ms$ scheme also has $UF - ESA$ security. The aggregate public key size and the final compressed signature size $\sigma$ is equal to the public key size and the signature size of a conventional signature, respectively. Also, $Razhi - ms$ only requires a matrix multiplication by a vector and a subtraction for signature verification. This issue makes $Razhi - ms$ more efficient than other lattice-based MS schemes.

**Author Contribution: Hamidreza Rahmati:** Conceptualization; Data curation; Formal analysis; Investigation; Methodology; Validation; Visualization; Writing - original draft. **Farhad Rahmati:** Conceptualization; Data curation; Project administration; Supervision; Validation; Writing - review & editing.

**Generative AI and AI-assisted technologies in the writing process:** During the preparation of this work, the authors used [ChatGPT] in order to [improve readability and language]. After using this tool/service, the authors reviewed and edited the content as needed and take full responsibility for the content of the publication.

**Funding:** This research did not receive any specific grant from funding agencies in the public, commercial, or not-for-profit sectors.

**Conflicts of interest:** The authors declare that they have no known competing financial interests or personal relationships that could have appeared to influence the work reported in this paper.